\documentstyle[referee]{mn}
\input epsf

\title{The ages and distances of globular clusters with the luminosity
function method: the case of M5 and M55.} 
\author[R. Jimenez \& P. Padoan]{Raul Jimenez,$^{1}$ and Paolo Padoan$^2$ \\
$^1$Royal Observatory, Blackford Hill, Edinburgh EH9-3HJ, UK \\
$^2$Theoretical Astrophysics Center, Juliane Maries Vej 30, 
DK-2100 Copenhagen 0, Denmark}

\begin{document}
\maketitle
\begin{abstract}
We present new age and distance determinations for the Galactic 
Globular Clusters M55 and M5, using the luminosity function method 
(Jimenez \& Padoan 1996, Padoan \& Jimenez 1997). 

We find an age of $11.8 \pm 1.5$ Gyr for M55 and $11.1 \pm 0.7$ Gyr for M5. 
This confirms previous results (Jimenez et al. 1996, Sandquist et al. 1996) and 
allows to conclude that {\bf the oldest stars in the Universe 
are not older than 14 Gyr}. 
 
We also find $m-M=14.13 \pm 0.11$ for M55, 
and $m-M=14.49  \pm 0.06$ for M5. These values agree with the
ones obtained using the tip of the red giant branch (Jimenez et al. 1996)
and the sub-dwarf fitting method (Sandquist et al. 1996). 

\end{abstract}


\section{Introduction}

An accurate determination of the ages and distances of Globular Clusters (GCs) 
is an important constraint for the age of the Universe, and for the theory of 
galaxy formation. In particular it is important to compute very accurate 
relative ages to understand if there is a spread in ages among the Galactic 
GCs or not.

The use of the stellar luminosity function (LF) to compute ages of GCs 
was first 
proposed by Paczynski (1984). Later on, Jimenez \& Padoan (1996) and 
Padoan \& Jimenez (1996) developed a method to determine {\it the age and 
the distance of a GC simultaneously}, using the LF. The method is described in 
detail in Padoan \& Jimenez (1996), where it is concluded, on the basis of 
artificial
data, that an uncertainty of about 0.6 Gyr in the age and 0.06 mag in the 
distance 
modulus can be achieved, if the number of stars, in 1 mag-wide luminosity bins,
is known with an uncertainty of 3\%. 

In this paper we use recent observations of the Galactic Globular Clusters 
M5 (Sandquist et al. 1996) and M55 (Desidera \& 
Ortolani, private communication) 
to apply the LF method and compute accurate ages and distance module.  
These two clusters are very adequate since M55 is a metal-poor one and M5 
has intermediate metallicity, so we can investigate the spread in ages 
(if any) in the formation of the GC system. 

In this letter we apply for the first time the LF method to real data and 
we show that the method is much more superior to traditional methods 
(isochrone fitting to the main sequence turn off point, $\Delta V$ method, 
or any other methods that involve the fitting of the main sequence turn off).
 The method is superior because it allows to determine the age and the 
distance simultaneously and independently and because the errors in 
computing the age and distance are straightforward to calculate. Furthermore, 
it gives age determinations with sufficient accuracy to make cosmological 
predictions.
This first application of our LF method to real data shows that our 
previous theoretical predictions were correct.

\section{The data}

In order to apply our LF method it is necessary to obtain the complete LF of
the globular cluster from almost the tip of the red giant branch (RGB), 
down to the
upper main sequence. The number of observed stars should be very large, 
in order
to keep statistical errors sufficiently low.  Recently, two LFs that 
fulfill these 
requirements have been obtained by Sandquist et al. (1996) for M5, and
by Desidera \& Ortolani (private communication) for M55.

M5 is a massive globular cluster, with an average metallicity
of $[Fe/H]=-1.17\pm0.01$, according to Sneden et al. (1992), and
$[Fe/H]=-1.4$, according to Zinn \& West (1984). 
Since it is  a high altitude cluster ($b=46.8^o$), 
it is not seriously affected by contamination and interstellar reddening. 
For the metallicity of this cluster we used $[Fe/H]=-1.3$, 
and adopt $Y=0.24$. The completeness of the LF is discussed in detail in 
Sandquist et al. (1996).   

M55 is among the poor-metal clusters. The main advantage of M55 is that 
it is not very concentrated and therefore it is possible to resolve its core into stars.
Again due to its high galactic latitude  ($b=-23^{o}$), interstellar reddening 
and contamination are negligible. We adopt a metallicity of $[Fe/H]=-1.9$
(Briley et al. 1993) and $Y=0.24$. The LF was provided to us by 
Desidera \& Ortolani (private communication.), who have performed extensive 
tests with artificial stars, in order to compute the completeness 
of the sample.

In this letter we use the data in the filter band $V$ for M55 and in $I$ for 
M5. This allows us to illustrate the robustness of the method in two very 
different bands. The other colours were used in both cases to remove the 
horizontal and asymptotic giant branch (AGB) stars. 
In the case of the horizontal branch it is quite easy to distinguish 
them from the RGB. Even if this is 
not the case for the AGB stars, the number of stars 
used in the LF method is so large in that bin that a small mistake in 
distinguishing RGB and AGB stars does not contribute significantly to 
the errors in the final 
age and distance determination.

In producing LFs to study both GCs we have used $[\alpha/Fe]=0.4$ and compute 
this effect in our solar-scaled tracks using the approach by Chieffi,  
 Straniero \& Salaris (1991). This value of alpha enhancement is well 
justified by spectroscopic observations of giants in GCs (Minniti et al 1996), 
and this enhancement is valid for a metallicity range 
($[Fe/H]=-1.5$ to $-2.0$).

\section{The luminosity function method}

To determine the age and the distance of a GC, two independent constraints 
are needed from the LF, that is to say the number of stars in two
different bins. One more bin is needed for the normalization, and a fourth
bin is useful to estimate the completeness of the data, but it is not 
used in this work, since the completeness was previously
estimated performing experiments with artificial stars (nevertheless, we 
have checked that the fourth bin gives the same completeness estimation as 
the artificial star tests).

We use a bin positioned at the RGB to normalize the LF, and two bins
around the sub-giant region to constrain age and distance modulus.
As discussed in Padoan \& Jimenez (1996), the precise position of the 
two bins at the sub-giant region is extremely important. In fact, our LF
method is based on the careful optimization of those two bins. 

The result of the optimization process is a contour plot of the quantity
$R(t,m-M)$, on the plane $(t,m-M)$, where:

\begin{eqnarray}
R^2(t,m-M)=[n_{2,\rm th}(t)-n_{2,\rm obs}(t,m-M)]^2 \nonumber \\
+[n_{3,\rm th}(t)-n_{3,\rm obs}(t,m-M)]^2   
\end{eqnarray}

where $t$ is the age, $m-M$ the distance modulus, $n_{i,\rm obs}$ the
normalized ratio $N_{i,\rm obs}/N_{1,\rm obs}$, $N_{i,\rm obs}$ the 
number of stars in the $i$th observational bin and $n_{i,\rm th}$ the 
corresponding theoretical ratio. The values of $n_{i,\rm th}$ are only 
functions of the age of the GC, since the shape of the theoretical LF
depends only on the age (for a given chemical composition), while
the values of $n_{i,\rm obs}$ are also functions of the distance modulus,
because the observational LF depends on the distance modulus, if the bins
are defined in absolute magnitudes.

If the set of bins 
is not optimal, the contour plot of $R$ shows only open lines,
which define a relation between age and distance modulus. Once the 
set of bins is optimized, the contour plot shows also closed lines, that define
both the age and the distance modulus of the GC at the same time.
If the lines start to become closed only for $R\le0.1$, the degeneracy
age-distance modulus is broken only if stellar counts are available
with uncertainty smaller than 10\%. This is the reason why the method
requires LF with a large number of stars and excellent photometry.

The optimization process, applied to M55 and M5 gave the following set of
bins:

\begin{eqnarray}
M_{V,optimal,M55}=(t_{Gyr}-9.0)\times0.05   \nonumber \\ 
+[4.01,3.01,2.01,0.01] mag \nonumber \\
M_{I,optimal,M5}=(t_{Gyr}-8.0)\times0.05  \nonumber \\ 
+[2.87,2.07,1.27,-3.13] mag \nonumber 
\end{eqnarray}

The optimal bins shift by $0.05$ mag/Gyr, as determined in our
previous work (Padoan \& Jimenez 1996). This is an essential point
in the attempt to obtain the contour plot of $R$.

Note that for M5 we could use two bins narrower than 1.0 mag, in
order to improve the sensitivity of the method, thanks to the very large 
number of stars in the bins and to the good quality of the photometry.

Our results for M55 is shown in Fig.~1, and for M5 in Fig.~2. In both 
cases, closed contour lines are obtained for $R\le 0.1$,
and the uncertainty in age and $m-M$, obtained at any given level
of $R$, is comparable for the two clusters, and similar to what we
previously predicted with artificial LFs (Padoan \& Jimenez, 1996)

Given the photometric uncertainty and the statistical uncertainty in the
stellar counts ($1/\sqrt{N}$, where N is the number of stars), we 
estimate a global uncertainty of 6\% for the case of M55, and of 
4\% for the case of M5. Entering Fig.~1 and Fig.~2 with $R=0.06$ and
$R=0.04$ respectively, one gets the results listed in Table~1.

\section{Discussion and Conclusions}

The results listed in Table~1 show that the age and the distance modulus 
of M55 are in good agreement with previous determinations by Mandushev et al. 
(1996) and Alcaino et al. (1992).

\begin{table}
\begin{center}
\begin{tabular}{ccc}
 & M5 & M55 \\
\hline\hline
age & $ 11.1 \pm 0.7$ & $11.8 \pm 1.5$ \\
m-M & $ 14.49 \pm 0.06$ & $14.13 \pm 0.11$ \\ 
\hline
\end{tabular}
\caption{The table gives the values for the age and distance modulus for 
M5 and M55. These values have been determined {\it simultaneously} using 
the luminosity function method described in the text.}
\end{center}
\end{table}

In the case of M5 the results agree with Sanquist et al. conclusions.
They estimate in fact an age of $13.5 \pm1$ Gyr, for $[Fe/H]=-1.17$,
and they state that the age would be 11.5 Gyr, for $[Fe/H]=-1.4$.
We use $[Fe/H]=-1.3$, and get an age of $11.1\pm0.7$ Gyr.   
They also estimate $m-M=14.50\pm0.07$ mag for $[Fe/H]=-1.17$,
and $m-M=14.41\pm0.07$ mag, for $[Fe/H]=-1.4$, using the sub-dwarf
fitting of the main sequence.
We get $m-M=14.49\pm0.06$ mag, for $[Fe/H]=-1.3$.
Note that in Padoan \& Jimenez (1996) we estimated a variation of 0.02 mag
in $m-M$, for a shift of 0.1 in metallicity; so we would predict 
$m-M=14.47\pm0.06$ mag, for $[Fe/H]=-1.4$.

\begin{figure}
\centering
\leavevmode
\epsfxsize=1.0
\columnwidth
\epsfbox{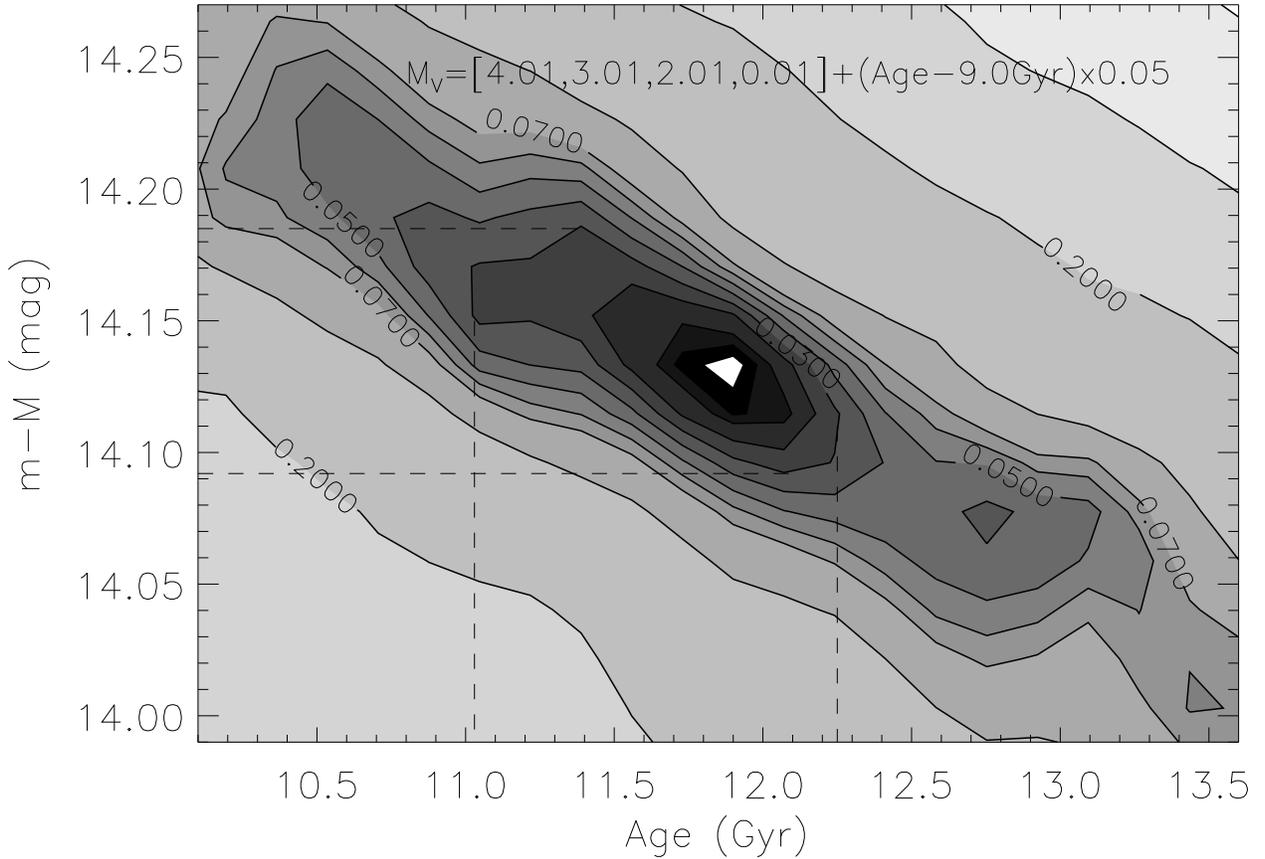}
\caption[]{The figure shows the contour plots of $R(t,m-M)$ (see text)  
 in
 determining simultaneously the distance modulus and age of M55. Notice that 
the contours closed around a central value, showing that the method works 
quite well in breaking the age-distance degeneracy.}
\end{figure}

The distance modulus measured with the LF method is therefore in 
excellent agreement with the distance modulus determined with the sub-dwarf
fitting. The uncertainty of our estimates is very small ($\pm0.06$ mag),
and no assumption on the age is required.

The LF method is superior to the main sequence turn-off (MSTO) method
(Chaboyer, Demarque \& Sarajedini 1996), to determine the absolute age of 
globular clusters, because it is not affected by the three largest sources of 
theoretical uncertainty affecting the MSTO method, that is to say the 
determination
of the value of the mixing length parameter, the morphology of the MSTO  
and the color-$T_{\rm eff}$ 
calibration (see Jimenez et al. 1996 for a detailed discussion of the 
main uncertainties in the MSTO method). Furthermore, the MSTO method 
needs to know the distance in order to determine the age, and it is 
unable to break this degeneracy.

The absolute ages, determined in this work for M55 and M5, seem to indicate 
that the oldest GCs are not older than 14 Gyr.

The LF method is a very powerful tool to investigate relative ages,
since most uncertainties of stellar evolution theories are in that case avoided.
From the comparison of the ages of M5 and M55 we can conclude that 
the age of the two GCs is not significantly different.

We conclude by remarking that most methods to determine age and distance
module of GCs share two common problems: some degree of dependence of 
age on distance modulus (or vice-versa), and a somewhat fuzzy procedure to
estimate the uncertainty of the final result. Our LF method, instead, gives
constraints for both age and distance modulus independently, and 
estimates both most probable values and uncertainties in a straightforward way.

\begin{figure}
\centering
\leavevmode
\epsfxsize=1.0
\columnwidth
\epsfbox{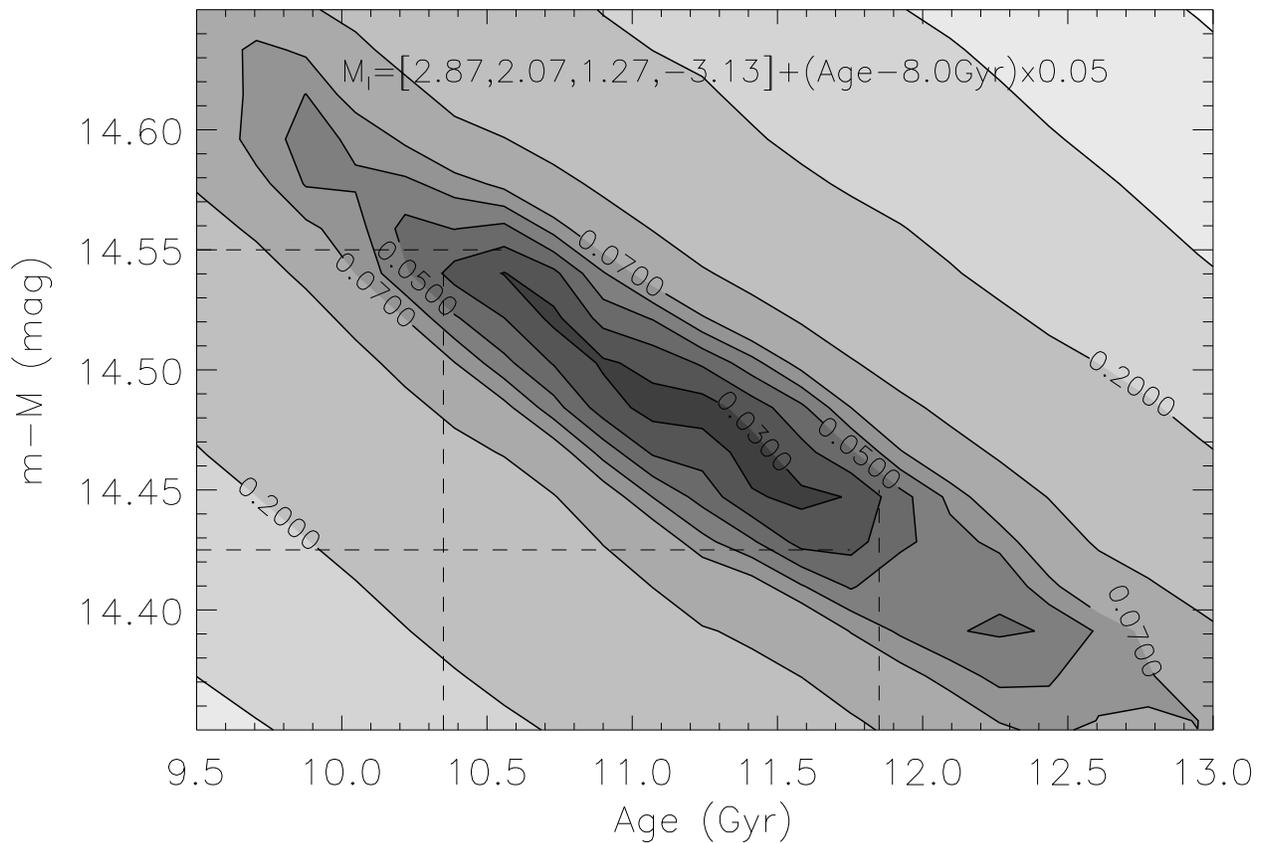}
\caption[]{The same as before but for M5. 
 The estimated uncertainty of 4\% is marked with dashed lines. } 
\end{figure}

On the basis of the present work, we think that very high quality
data for GCs, together with the LF method, may shed new light
on the problems of the age of the oldest stars in the Universe and the 
formation of the Galaxy.

\section*{acknowledgements}
We are grateful to S. Desidera \% S. Ortolani for providing us with 
unpublished data of M55.

\end{document}